\documentclass[aps,prb,reprint,nofootinbib,superscriptaddress,twocolumn,showpacs,showkeys,longbibliography,amsmath,amssymb]{revtex4-1}
\usepackage{graphicx}
\usepackage{dcolumn}
\usepackage{bm}
\usepackage{braket}
\usepackage{subfigure}
\usepackage[colorlinks,bookmarks=false,citecolor=blue,linkcolor=red,urlcolor=blue]{hyperref}
\usepackage[english]{babel}
\usepackage{changes}

\begin{document}

\title{Thermal Hall conductivity with sign change in the Heisenberg-Kitaev kagome magnet}
\author{Kangkang Li}
\email{physeeks@163.com}
\affiliation{Department of Physics, Zhejiang Normal University, Jinhua, 321004, China}

\begin{abstract}
The Heisenberg-Kitaev (HK) model on various lattices have attracted a lot of attention because they may lead to exotic states such as quantum spin liquid and topological orders. The rare-earth-based kagome lattice (KL) compounds $\rm{Mg_2RE_3Sb_3O_{14}}$ $\rm{(RE=Gd, Er)}$ and $\rm{(RE=Nd)}$ have ${\bf q=0}$, $120^\circ$ order and canted ferromagnetic (CFM) order, respectively. Interestingly, the HK model on the KL has the same ground state long-range orders. In the theoretical phase diagram, the CFM phase resides in a continuous parameter region and there is no phase change across special parameter points, such as the Kitaev ferromagnetic (KFM) point, the ferromagnetic (FM) point and its dual FM point. However, a ground state property cannot distinguish a system with or without topological nontrivial excitations and related phase transitions. Here, we study the topological magnon excitations and related thermal Hall conductivity in the HK model on the KL with CFM order. The CFM phase can be divided into two regions related by the Klein duality, with the self dual KFM point as their boundary. We find that the scalar spin chirality which is intrinsic in the CFM order changes sign across the KFM point. This leads to the opposite Chern numbers of corresponding magnon bands in the two regions, and also the sign change of the magnon thermal Hall conductivity.
\end{abstract}
\maketitle

\section{Introduction}\label{sectoin1}
The Heisenberg-Kitaev (HK) model on various lattices such as triangular, kagome, pyrochlore, hyperkagome and fcc lattices have attracted a lot of attention, because there are frustrations from the competing exchange couplings coexisting with the geometric frustration of the underlying lattices, which may leads to exotic states such as quantum spin liquid and topological orders\cite{Kitaev,Rou,Lee,Nasu,Taka,Yao,Karga,Jah,Jah2,Kishi,Kim}. The studies were also connected to real materials, such as the layered honeycomb materials $\alpha$-$\rm{RuCl_3}$, $\rm{Na_2IrO_3}$ and $\alpha$-$\rm{Ir_2IrO_3}$, the three-dimensional (3D) honeycomb material $(\beta$-$\gamma)$-$\rm{Li_2IrO_3}$, and the iridium oxides with triangular lattices\cite{Cha,Jiang,Reu,Kim2,Scha,Sing,Choi,Reu2,Cha2,Oka,Price,Sela,Rau,
Yama,Siz,Cha3,Oku,Got,Dey,Beck,Cat,Jack,Rou2,Lee2,Plum,Kubota,Kim4,John,Bast,Zhu,Goh}.

The rare-earth-based kagome lattice (KL) compounds $\rm{Mg_2RE_3Sb_3O_{14}}$ $\rm{(RE=Gd, Er)}$ and $\rm{(RE=Nd)}$ have ${\bf q=0}$, $120^\circ$ order and canted ferromagnetic (CFM) order\cite{Dun,Sche}, respectively. These compounds, except for $\rm{RE=Gd}$, have an effective spin with $S=1/2$ on the KL\cite{Dun2}. Since the exchange couplings between the nearest-neighbor (NN) spins are anisotropic, Morita {\it et al} studied the ground state of the classical and quantum spin HK model with anisotropic bond-dependent Kitaev interactions on the KL\cite{Morita}. Interestingly, the ground state has the same type ${\bf q=0}$, $120^\circ$ order and CFM order as those observed in the compounds. In the theoretical phase diagram, the CFM order resides in a continuous parameter region, and there is no phase change across special parameter points, such as the Kitaev ferromagnetic (KFM) point, the ferromagnetic (FM) point and its dual FM point. However, a ground state property cannot distinguish a system with or without topological nontrivial excitations and related phase transitions. The non-coplanar CFM order naturally has nonzero scalar spin chirality, which will provide a vector potential for magnons\cite{Owerre}. Then one may wonder are there topological magnon excitations and related thermal Hall effect in the system. Furthermore, is it possible for a topological phase transition of magnon excitations to happen on the continuous CFM ground state.

\begin{figure}[bth]
      \centering
      \includegraphics[width=0.48\textwidth]{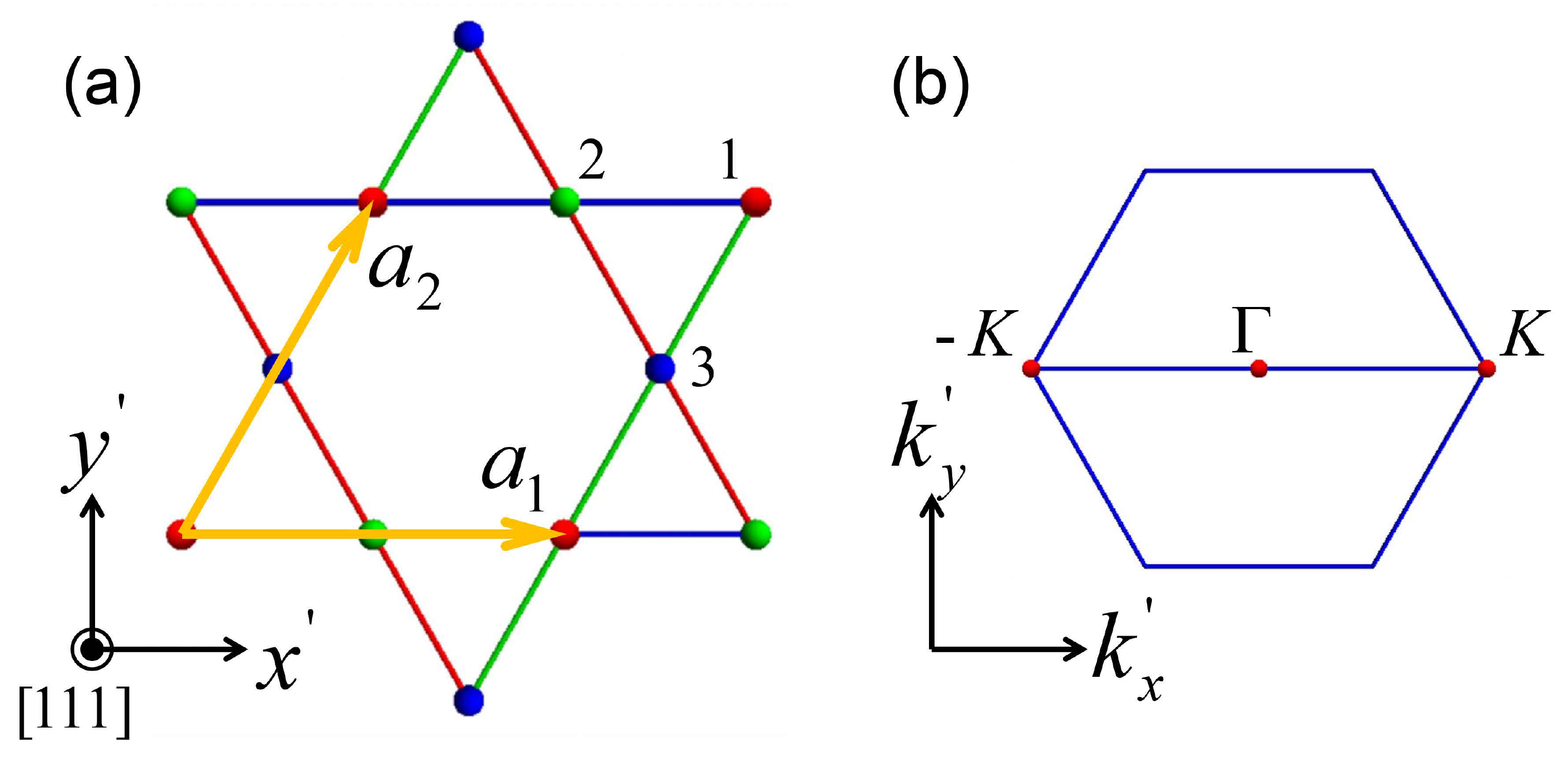}
      \caption{(a) The structure of the KL. There are three spins reside in a primitive cell, which are denoted by red, green and blue sites. Red, green and blue bonds between NN sites $(i,j)$ carry three distinct Kitaev couplings $S_i^xS_j^x$, $S_i^yS_j^y$ and $S_i^xS_j^x$, respectively. Meanwhile, there are isotropic Heisenberg couplings in all the NN bonds. The KL sits on the (111) plane, and we define a new 2D frame $x^\prime y^\prime$ on the KL with basis vectors $\bm{a}_1=(1, 0)$ and $\bm{a}_2=(1,\sqrt{3})/2$. (b) The first Brillouin zone of the KL.}
      \label{Fig1}
\end{figure}

\begin{figure*}[tbh]
      \centering
      \includegraphics[width=0.9\textwidth]{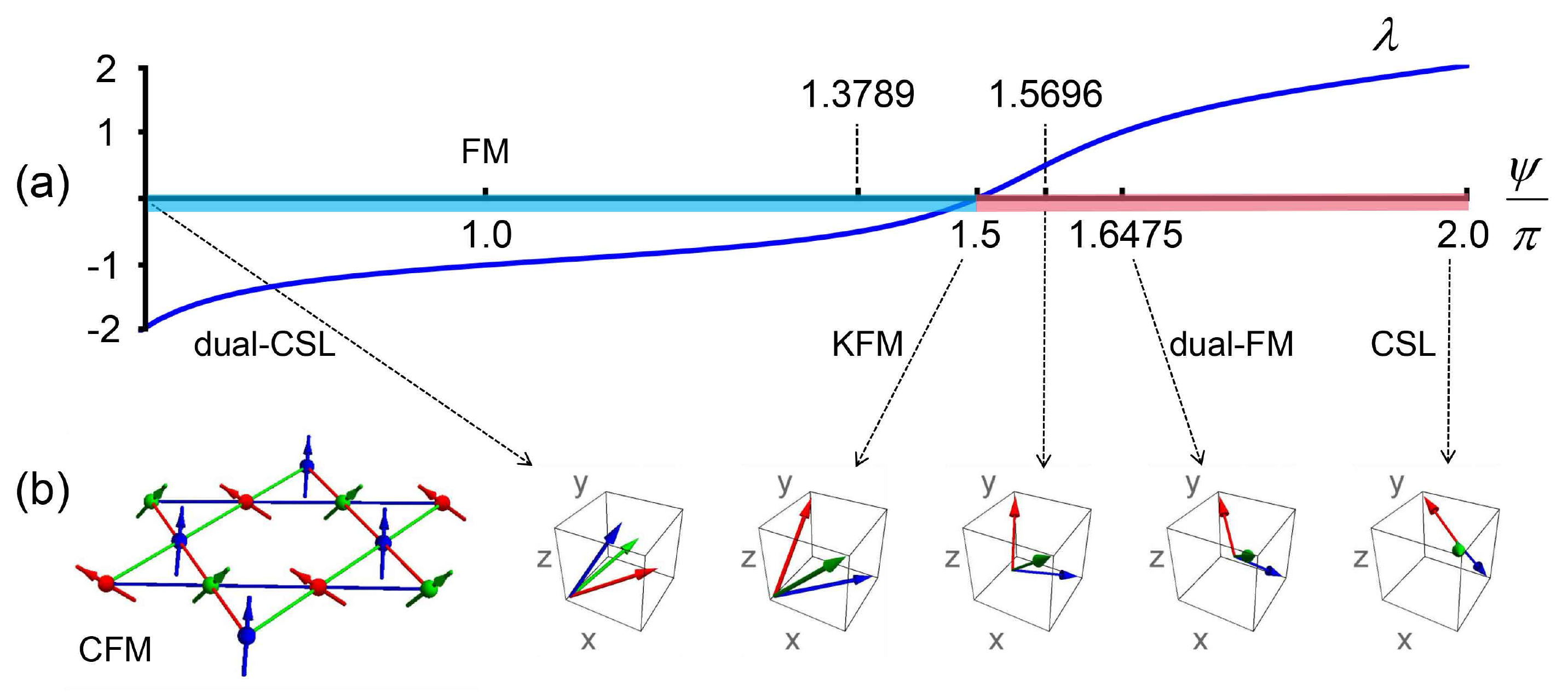}
      \caption{(a) Evolution of $\lambda$ [Eq. (\ref{lambda})] with $\psi$ and the phase diagram of topological magnons in the CFM phase. In the region $\psi\in(\pi-\arctan(2),1.5\pi)$ shaded cyan, the three magnon bands have Chern numbers 1, 0, -1, wihle in the region $\psi\in(1.5\pi,2\pi)$ shaded pink they are -1, 0, 1. This denotes a topological phase transition at the KFM point. Note that the Chern numbers vanish at the FM, dual FM, CSL, dual CSL and KFM points where the magnon bands are gapless. The points $\psi=1.3789\pi$ and $\psi=1.5696\pi$ are a normal pair of dual points. (b) Schematic CFM spin order with $\chi=\eta=\zeta=1$ and the evolution of the directions of the three spins in a primitive cell with $\psi$.}
      \label{Fig2}
\end{figure*}

Here, we study the topological magnon excitations and related thermal Hall conductivity in the HK model on the KL with CFM order. In the CFM order, the solid angle spanned by the three spins in a unit cell changes continuously with the model parameter\cite{Yang}. However, the continuous CFM phase can be divided into two regions related by the Klein duality, with the self dual KFM point as their boundary\cite{Yang}, and we are surprised to find that the scalar spin chirality which is intrinsic in the CFM order changes sign across the KFM point. This leads to the opposite Chern numbers of corresponding magnon bands in the two regions, and also the sign change of the magnon thermal Hall conductivity.

This paper is organized as follows. In section \ref{sectoin2}, we revisit the CFM phase in the model and investigate the scalar spin chirality. In section \ref{sectoin3}, we present the magnon band structures and discuss their topological properties. In section \ref{sectoin4}, we show the transverse thermal Hall conductivity with sign change phenomena. Finally, a summary is given in section \ref{sectoin5}.

\section{Model, the CFM phase and scalar spin chirality}\label{sectoin2}
We first revisit the model and the CFM phase\cite{Morita,Yang}. Consider interacting spins reside on the KL as shown in Fig. \ref{Fig1}(a). The HK model is described by the spin Hamiltonian
\begin{align}
H=\sum_{\langle ij\rangle}(J{\bf S}_i\cdot{\bf S}_j+KS_i^{\gamma_{ij}}S_j^{\gamma_{ij}}),
\label{model}
\end{align}
where ${\bf S}_i$ and ${\bf S}_j$ are spin $S=1/2$ spins reside on the NN lattice sites, and $J$ and $K$ denote the Heisenberg and Kitaev exchange couplings, respectively. The Cartesian components $\gamma_{ij}$ equals $x$, $y$ or $z$, depending on the bond type as shown in Fig. \ref{Fig1}(a). The model can be parameterized by
\begin{align}
J=\cos\psi,\quad K=\sin\psi,\quad \psi \in [0, 2\pi),
\end{align}
with the energy unit $J^2+K^2=1$.

In the classical ground state phase diagram of the HK model on KL, there is a long-range ordered CFM phase for $\psi\in[\pi-\arctan(2),2\pi]$, which has eightfold degeneracy. At the phase boundary are the classical spin liquid (CSL) state and dual CSL state corresponding to the Heisenberg point with $K=0$ and its dual point, respectively. Inside the CFM phase, there are other three special points, the KFM point with $J=0$, the FM point with $K=0$ and its dual FM point (see Fig. \ref{Fig2}(a)). There is no phase transition of the ground state across the whole CFM phase. However, we find that the collective magnon excitations on the CFM ground state do have topological phase transition at the KFM point as shown in Fig. \ref{Fig2}(a). Interestingly, we can divide the CFM phase into two regions related by the Klein duality\cite{Kimchi,Morita,Yang}, with the self-dual KFM point as the boundary. The Klein duality preserve the form of the Hamiltonian but alter the value of the parameters $J$ and $K$, and in the KL it reads
\begin{align}
\widetilde{H}=\sum_{\langle ij\rangle}(\widetilde{J}\,\widetilde{{\bf S}}_i\cdot\widetilde{{\bf S}}_j+\widetilde{K}\widetilde{S}_i^{\gamma_{ij}}\widetilde{S}_j^{\gamma_{ij}}),
\end{align}
where
\begin{align}
\widetilde{J}=-J, \quad \widetilde{K}=2J+K,
\end{align}
and the spin operators transform to
\begin{align}
\widetilde{{\bf S}}_1=&\left(
\begin{array}{c}
-S_1^x\\[6pt]
S_1^y\\[6pt]
-S_1^z
\end{array}
\right),
\widetilde{{\bf S}}_2=\left(
\begin{array}{c}
S_2^x\\[6pt]
-S_2^y\\[6pt]
-S_2^z
\end{array}
\right),\notag\\
\widetilde{{\bf S}}_3=&\left(
\begin{array}{c}
S_3^x\\[6pt]
S_3^y\\[6pt]
S_3^z
\end{array}
\right).
\end{align}

\begin{figure}[tbh]
      \centering
      \includegraphics[width=0.48\textwidth]{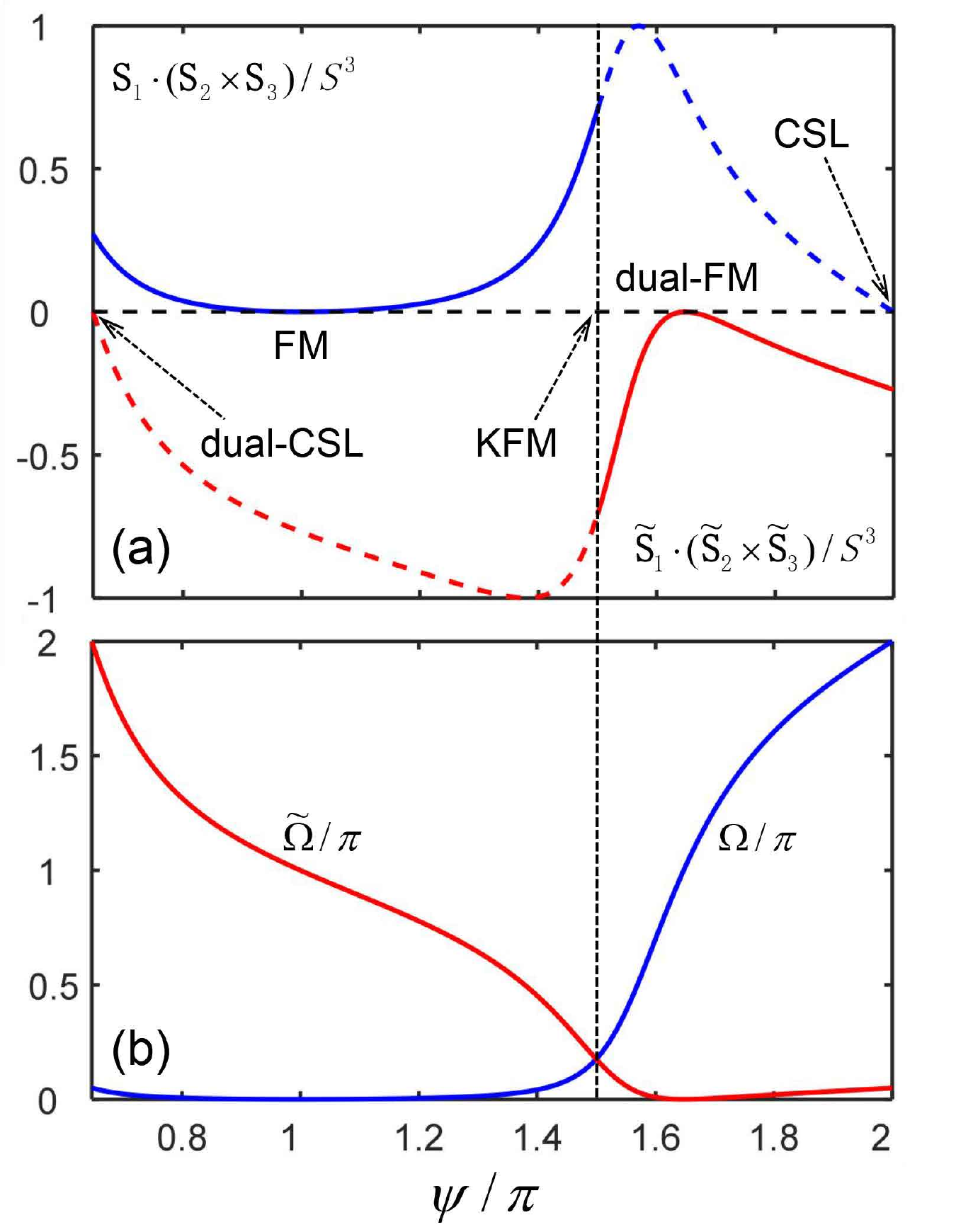}
      \caption{(a) Evolutions of the scalar spin chirality [Eq. (\ref{scalar})] (blue line) and its dual one [Eq. (\ref{dualscalar})] (red line) with $\psi$. In the left side of the KFM point, the scalar spin chirality for magnons evolves with the blue line, while in the right side, it jumps to the red line. (b) Evolutions of the solid angle $\Omega$ and its dual one $\widetilde{\Omega}$ with $\psi$. In the left side of the KFM point, the solid angle $\Omega$ spanned by spins ${\bf S}_{1,2,3}$ is much smaller, while the situation is opposite in the right side.}
      \label{Fig3}
\end{figure}

The spin structure of the CFM phase can be described by the directions of the three spins in a primitive cell\cite{Yang}
\begin{align}
{\bf S}_1=&\xi S\left(
\begin{array}{c}
-\lambda\chi\\[6pt]
\eta\\[6pt]
\zeta
\end{array}
\right),
{\bf S}_2=\xi S\left(
\begin{array}{c}
\chi\\[6pt]
-\lambda\eta\\[6pt]
\zeta
\end{array}
\right),\notag\\
{\bf S}_3=&\xi S\left(
\begin{array}{c}
\chi\\[6pt]
\eta\\[6pt]
-\lambda\zeta
\end{array}
\right),
\end{align}
where $\xi=1/\sqrt{2+\lambda^2}$ and
\begin{align}
\lambda=(\cos\psi+\sin\psi+\sqrt{5+4\cos2\psi+\sin2\psi})/(2\cos\psi).
\label{lambda}
\end{align}
The eight degenerate spin orders is given by $\chi=\pm1$, $\eta=\pm1$ and $\zeta=\pm1$.
The magnon excitations of the eight different spin orders have the same band structures, however different wave functions and therefore different topological properties. Here, we study the order given by $\chi=\eta=\zeta=1$, which is the same as the one observed in experiment\cite{Sche}, as shown in Fig. \ref{Fig2}(b).

Since the three spins in a primitive cell are non-coplanar in the CFM order, there is intrinsically non-zero scalar spin chirality. In the spin order with $\chi=\eta=\zeta=1$, the scalar spin chirality is
\begin{align}
{\bf S}_1\cdot({\bf S}_2\times{\bf S}_3)/S^3=\frac{-\lambda^3+3\lambda+2}{(\lambda^2+2)^{3/2}},
\label{scalar}
\end{align}
and its dual one is
\begin{align}
\widetilde{{\bf S}}_1\cdot(\widetilde{{\bf S}}_2\times\widetilde{{\bf S}}_3)/S^3=\frac{-\lambda^3+3\lambda-2}{(\lambda^2+2)^{3/2}}.
\label{dualscalar}
\end{align}
It turns out that the chirality is always positive except at the FM and CSL points, where it vanishes. Oppositely, the dual chirality is always negative and vanishes at the dual FM and dual CSL points, as shown in Fig. \ref{Fig3}(a). Here we are surprised to find that the scalar spin chirality for the magnons changes sign at the KFM point, where it jumps from the original chirality to its dual one. We attribute this to the relative size of the corresponding solid angle and its dual one. In the left side of the KFM point, the solid angle $\Omega$ spanned by spins ${\bf S}_{1,2,3}$ is smaller than its dual one $\widetilde{\Omega}$ (see Fig. \ref{Fig3}(b)), which is spanned by the dual spins $\widetilde{\bf S}_{1,2,3}$, and the chirality for magnons is ${\bf S}_1\cdot({\bf S}_2\times{\bf S}_3)$. However, the situation is opposite in the right side of the KFM point, and now the chirality for magnons will be $\widetilde{{\bf S}}_1\cdot(\widetilde{{\bf S}}_2\times\widetilde{{\bf S}}_3)$.

 The scalar spin chirality can provides a vector potential for magnons, which will lead to topological magnons and thermal Hall effect, as previous studies on ferromagnets with Dzyaloshinskii-Moriya (DM) interaction induced vector potential\cite{Katsura,Ono,Hir,Chis,Owerre2,Cao}. The more interesting thing here is the scalar spin chirality changes sign across the KFM point. This will lead to the opposite Chern numbers of corresponding magnon bands in the two dual regions, and also the sign change of the magnon thermal Hall conductivity, as we show in the next sections.

 \begin{figure*}[tbh]
      \centering
      \includegraphics[width=0.9\textwidth]{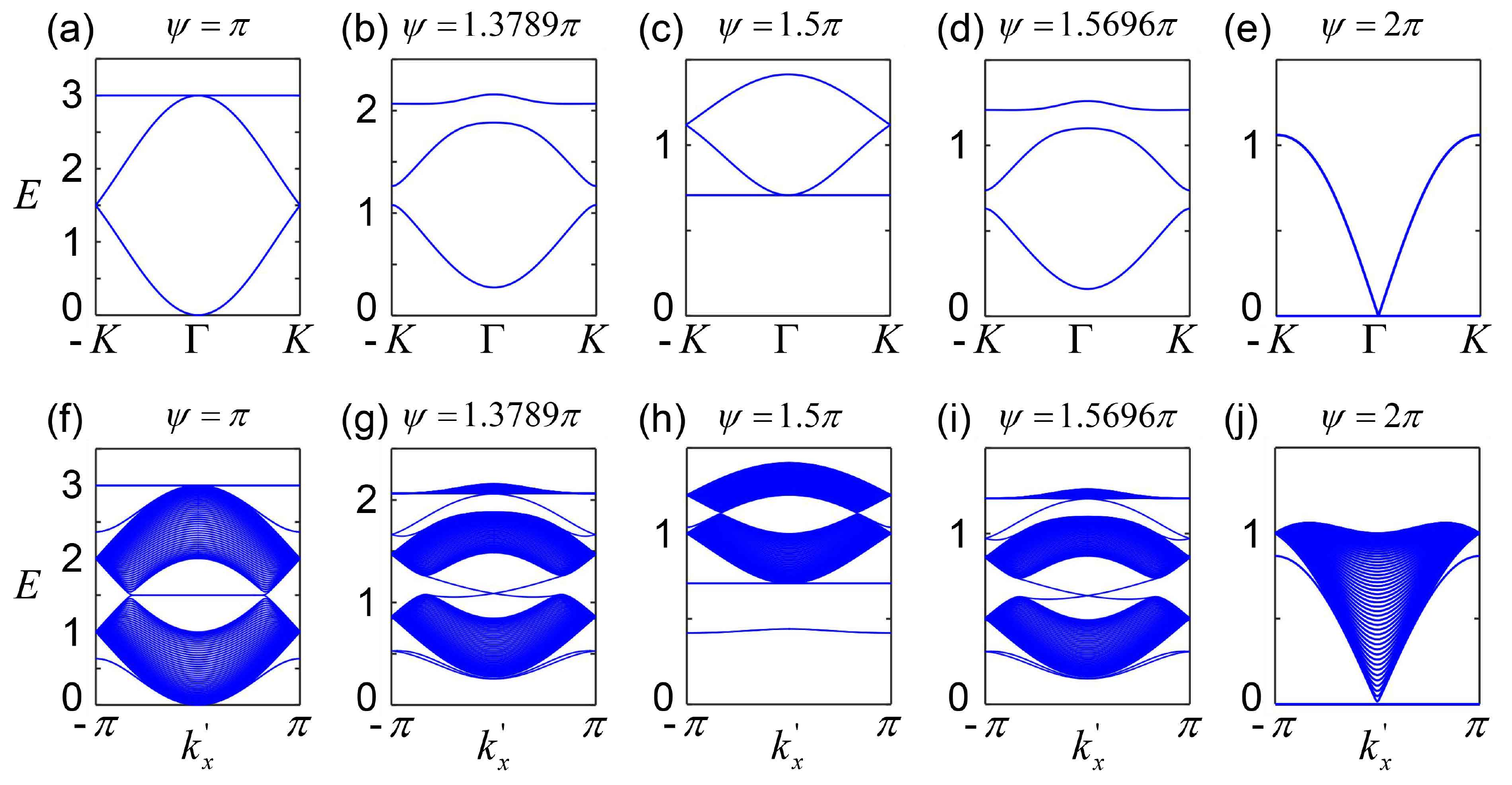}
      \caption{Top panel: magnon bands of the CFM order at the (a) FM point, (c) KFM point, (e) CSL point, and (b) and (d) a pair of dual points. The magnon bands are gapped in all the CFM phase, except at the CSL, dual CSL, FM, dual FM and KFM points. The points (b) $\psi=1.3789\pi$ and (d) $\psi=1.5696\pi$ also denoted in Fig. \ref{Fig2} (a) are a pair of dual points, and their band structures are the same except for the energy scale. The bands in (b) carry Chern numbers 1, 0, -1, while they are -1, 0, 1 in (d). Bottom panel: the corresponding magnon bands in a strip geometry. There are in-gap edge modes for the gapped band structures, which are promised by the nonzero Chern numbers of the corresponding bulk bands.}
      \label{Fig4}
\end{figure*}

\section{Band structure and topological magnons}\label{sectoin3}
Now we turn to study the collective magnon excitations on the CFM order with $\chi=\eta=\zeta=1$. Starting from the Hamiltonian (\ref{model}), after the Holstein-Primakoff (HP)\cite{HP} and Fourier transformations, we get the magnon Hamiltonian matrix $h(\bm{k}^{\prime})$ (see Appendix). Then after diagonalizing $I_-h(\bm{k}^{\prime})$, we get three magnon bands. The magnon bands are gapped in all the CFM phase, except at the CSL, dual CSL, FM, dual FM and KFM points as shown in the top panel of Fig. \ref{Fig4}. Due to the same form of the Hamiltonian, the band structures are the same for any pair of dual points, except for the energy scale. At the FM and dual FM points, the magnon bands are the same with the isotropic Heisenberg model on KL\cite{Chis}. At the CSL and dual CSL points, there are zero energy flat bands, which means the magnon modes now can excite without energy cost. Moreover, the possibilities for the excited magnon modes with different $\bm{k}^{\prime}$ are the same, and this will lead to the breakdown of the long-range order. Here the breakdown of the CFM order gives the CSL phase. At the KFM point there is also a flat band, however with finite energy. Hence, to excite the flat band modes now will cost energy, and the CFM order is protected by the energy gap.

Though the band structures of the dual points are the same, the wave functions and topological properties are not necessarily the same. To show the topological property of the gapped magnon bands, we calculate their Chern numbers. The Chern number of the $n$th band is defined by the integration of the Berry curvature over the first Brillouin zone\cite{Xiao}
\begin{align}
C_n=\frac{1}{2\pi}\int_{\rm BZ}{\rm d}k_x^{\prime}{\rm d}k_y^{\prime}B_{k_x^{\prime}k_y^{\prime}}^n,
\end{align}
with
\begin{align}
B_{k_x^{\prime}k_y^{\prime}}^n\!\!={\rm i}\!\!\sum_{n^\prime\ne n}\!\!\frac{\langle\phi_n|\frac{\partial h(\bm{k}^{\prime})}{\partial k_x^{\prime}}|\phi_{n^\prime}\rangle\langle\phi_{n^\prime}|\frac{\partial h(\bm{k}^{\prime})}{\partial k_y^{\prime}}|\phi_n\rangle\!\!-\!\!(k_x^{\prime}\leftrightarrow k_y^{\prime})}{(E_n-E_{n^\prime})^2},
\end{align}
where $E_n$ and $\phi_n$ are the eigenvalue and eigenvector of the $n$th band respectively. It turns out that the Chern numbers are $(C_1, C_2, C_3)=(1, 0, -1)$ in the region $\psi\in(\pi-\arctan(2),1.5\pi)$, while they are $(-1, 0, 1)$ in the region $\psi\in(1.5\pi, 2\pi)$. This denotes a topological phase transition at the KFM point $\psi=1.5\pi$ (see Fig. \ref{Fig2}(a)). We show the gapped magnon bands of a pair of dual points in Fig. \ref{Fig4}(b) and \ref{Fig4}(d), and their Chern numbers are opposite. As mentioned in the above section, it is the sign change of the scalar spin chirality for magnons that leads to the opposite Berry curvatures and Chern numbers of corresponding magnon bands in the two regions related by Klein duality. Due to the bulk-edge correspondence\cite{Hatsu}, the nonzero Chern numbers will promise in-gap edge modes in a strip geometry as shown in Fig. \ref{Fig4}(g) and \ref{Fig4}(i).

\section{Transverse thermal Hall conductivity with sign change}\label{sectoin4}

\begin{figure*}[tbh]
      \centering
      \includegraphics[width=0.9\textwidth]{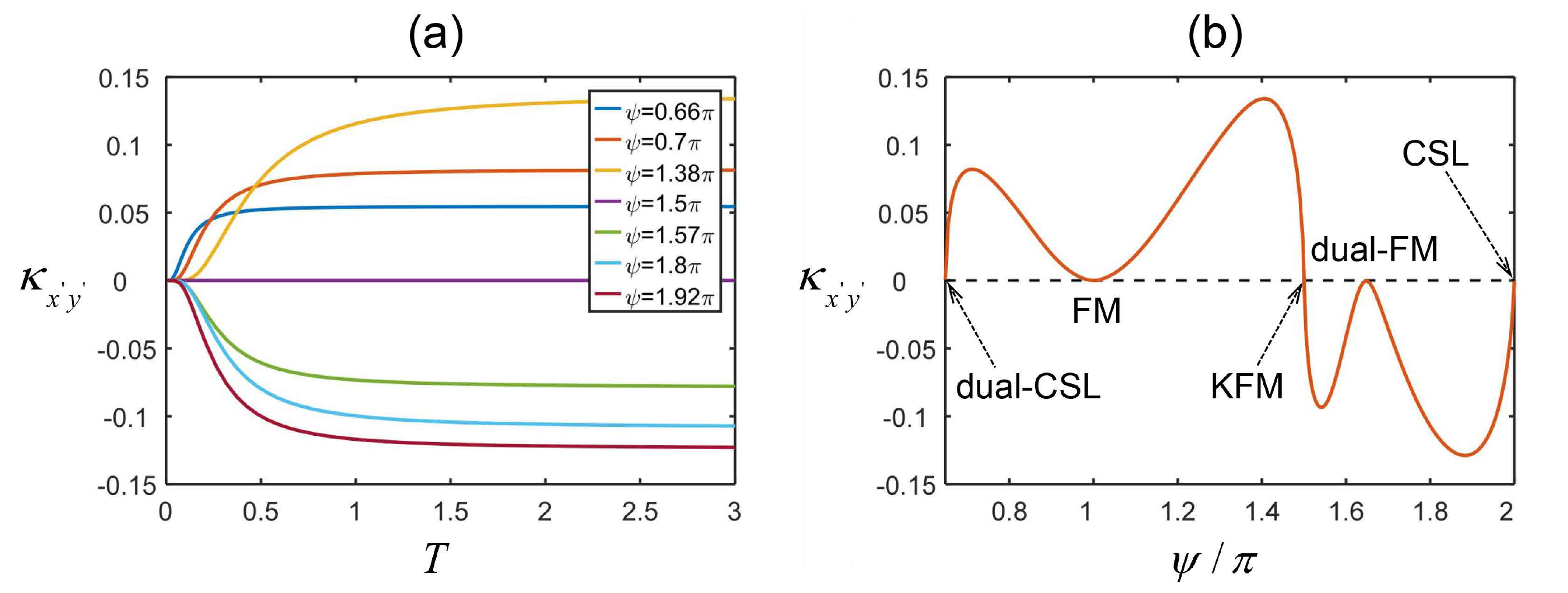}
      \caption{(a) The transverse thermal conductivity as function of temperature for different $\psi$. For $\psi\in(\pi-\arctan(2), 1.5\pi)$, the conductivity is always positive, while for $\psi\in(1.5\pi, 2\pi)$ it is always negative. At the CSL, dual CSL, FM, dual FM and KFM points, the conductivity is always zero. (b) The transverse thermal conductivity as function of $\psi$ with $T=3$. There is sign change at the KFM point. Here we set $\hbar=k_B=1$.}
      \label{Fig5}
\end{figure*}

As previous studies on ferromagnets with DM interaction induced vector potential, the scalar spin chirality here can also provides a vector potential for magnons, which will lead to the magnon thermal Hall effect. Furthermore, the sign change of the chirality here will also induce the sign change of the thermal conductivity simultaneously. The transverse thermal Hall conductivity can be calculated as\cite{Matsu,Matsu2,Matsu3}
\begin{align}
\kappa_{x^{\prime}y^{\prime}}=\frac{k_B^2T}{(2\pi)^2\hbar}\sum_n\int_{\rm BZ}c_2(\rho_n)B_{k_x^{\prime}k_y^{\prime}}^n{\rm d}k_x^{\prime}{\rm d}k_y^{\prime},
\end{align}
with the sum running over the three bands and the integral is over the first Brillouin zone. $\rho_n=1/(\exp(E_n/k_BT)-1)$ is the Bose distribution with $E_n$ as the $n$th eigenvalue. $c_2$ is given by
\begin{align}
c_2(\rho_n)=(1+\rho_n)(\rm{ln}\frac{1+\rho_n}{\rho_n})^2-(\rm{ln}\rho_n)^2-2Li_2(-\rho_n),
\end{align}
where $\rm{Li}_2(x)$ is the polylogarithm function of order 2.

As shown in Fig. \ref{Fig5}(a) and \ref{Fig5}(b), the transverse thermal conductivity for $\psi\in(\pi-\arctan(2), 1.5\pi)$ is always positive, while it is always negative for $\psi\in(1.5\pi, 2\pi)$. At the CSL, dual CSL, FM, dual FM and KFM points, the magnon bands are gapless, leading to vanishing thermal Hall conductivity. There is sign change of the thermal conductivity at the KFM point, and we attribute it to the sign change of the scalar spin chirality. As the chirality changes sign, the vector potential for magnons becomes opposite, and the propagating magnons will be deflected in the opposite direction, which results in the opposite transverse conductivity.

The sign change of the thermal Hall conductivity is also consistent with the opposite Chern numbers of corresponding magnon bands. Consider the high temperature limit of the transverse thermal conductivity\cite{Mook,Cao}
\begin{align}
\kappa_{x^{\prime}y^{\prime}}^{\rm lim}=\lim_{T\rightarrow \infty}\kappa_{x^{\prime}y^{\prime}}=-\frac{k_B}{(2\pi)^2\hbar}\sum_n\int_{\rm BZ}E_nB_{k_x^{\prime}k_y^{\prime}}^n{\rm d}k_x^{\prime}{\rm d}k_y^{\prime},
\end{align}
it can be further simplified as
\begin{align}
\kappa_{x^{\prime}y^{\prime}}^{\rm lim}\propto-\frac{k_B}{(2\pi)^2\hbar}\sum_nC_n\overline{E}_n,
\end{align}
where the $\bm{k}^{\prime}$-dependent band energy is replaced by the average $\overline{E}_n$. Then the high temperature thermal conductivity of the phase with Chern numbers $(1, 0, -1)$ is $\overline{E}_3-\overline{E}_1$ which is positive, while the conductivity of the phase with Chern numbers $(-1, 0, 1)$ is $\overline{E}_1-\overline{E}_3$ which is negative. Then the sign change of the chirality, the opposite Chern numbers of corresponding bands and the sign change of the thermal conductivity are consistent with each other.

In the end of this section, we will discuss briefly of the eight degenerate CFM spin orders. We find that the chirality of the orders with $[\chi \eta \zeta]=[111]$, $[1\bar{1}\bar{1}]$, $[\bar{1}1\bar{1}]$, and $[\bar{1}\bar{1}1]$ is
\begin{align}
{\bf S}_1\cdot({\bf S}_2\times{\bf S}_3)/S^3=\frac{-\lambda^3+3\lambda+2}{(\lambda^2+2)^{3/2}},
\end{align}
and its dual one is
\begin{align}
\widetilde{{\bf S}}_1\cdot(\widetilde{{\bf S}}_2\times\widetilde{{\bf S}}_3)/S^3=\frac{-\lambda^3+3\lambda-2}{(\lambda^2+2)^{3/2}}.
\end{align}
However, the chirality of the orders $[\bar{1}\bar{1}\bar{1}]$, $[\bar{1}11]$, $[1\bar{1}1]$, and $[11\bar{1}]$ are opposite
\begin{align}
{\bf S}_1\cdot({\bf S}_2\times{\bf S}_3)/S^3=\frac{\lambda^3-3\lambda-2}{(\lambda^2+2)^{3/2}},
\end{align}

\begin{align}
\widetilde{{\bf S}}_1\cdot(\widetilde{{\bf S}}_2\times\widetilde{{\bf S}}_3)/S^3=\frac{\lambda^3-3\lambda+2}{(\lambda^2+2)^{3/2}}.
\end{align}
Besides, the solid angles $\Omega$ and $\widetilde{\Omega}$ keep the same for all the eight orders. This suggests that the Chern numbers of the magnon bands and the sign of the thermal Hall conductivity are the same for the orders $[111]$, $[1\bar{1}\bar{1}]$, $[\bar{1}1\bar{1}]$, and $[\bar{1}\bar{1}1]$, while they are opposite to that of the orders $[\bar{1}\bar{1}\bar{1}]$, $[\bar{1}11]$, $[1\bar{1}1]$, and $[11\bar{1}]$.

\section{Summary}\label{sectoin5}
We have studied the topological magnon excitations and related thermal Hall conductivity in the HK model on the KL with CFM order. The CFM phase can be divided into two regions related by the Klein duality\cite{Kimchi,Morita,Yang}, with the self dual KFM point as their boundary. We find that the scalar spin chirality which is intrinsic in the CFM order changes sign across the KFM point. This leads to the opposite Chern numbers of corresponding magnon bands in the two regions, and also the sign change of the magnon thermal Hall conductivity. Moreover, we have checked that for the coplanar ${\bf q=0}$, $120^\circ$ order\cite{Morita} the magnon bands are always gapless, and there is no thermal Hall conductivity. Interestingly, the rare-earth-based KL compounds $\rm{Mg_2RE_3Sb_3O_{14}}$ $\rm{(RE=Gd, Er)}$ and $\rm{(RE=Nd)}$ have the same ${\bf q=0}$, $120^\circ$ order and CFM order\cite{Dun,Sche}, respectively. Though the chiral edge modes are difficult to measure in experiment, the sign structure of the thermal conductivity is more accessible\cite{Yoko}. Therefore, the study of the topological magnons and related thermal Hall conductivity here will contribute to the understanding of related compounds.

\section*{Acknowledgements}
We thank Changle Liu for helpful discussions. This work was supported by the National Natural
Science Foundation of China (Grant NO. 12104407) and the Natural Science Foundation of Zhejiang Province (Grant NO. LQ20A040004).

\section*{Appendix: magnon Hamiltonian matrix}
We denote the directions of spins ${\bf S}_{i=1,2,3}$ by their polar angles $\theta_{1,2,3}$ and azimuthal angles $\phi_{1,2,3}$ in the global frame. The HP transformation for a spin in its local frame reads
\begin{align}
S_x^0=&\frac{\sqrt{2S}}{2}(a+a^\dagger),\\
S_y^0=&\frac{\sqrt{2S}}{2{\rm i}}(a-a^\dagger),\\
S_z^0=&S-a^\dagger a,
\end{align}
where $a^\dagger$ and $a$ are the magnon creation and annihilation operators respectively, which obey the boson commutation rules.
Then by multiplying a rotation matrix we get the HP transformation for spin ${\bf S}_i$ in the global frame
\begin{align}
\left(
\begin{array}{c}
S_{ix}\\[8pt]
S_{iy}\\[8pt]
S_{iz}
\end{array}
\right)=\left(
\begin{array}{ccc}
\cos\theta_i\cos\phi_i & -\sin\phi_i & \sin\theta_i\cos\phi_i\\[8pt]
\cos\theta_i\sin\phi_i & \cos\phi_i & \sin\theta_i\sin\phi_i\\[8pt]
-\sin\theta_i & 0 & \cos\theta_i
\end{array}
\right)
\left(
\begin{array}{c}
S_{ix}^0\\[8pt]
S_{iy}^0\\[8pt]
S_{iz}^0
\end{array}
\right).
\end{align}
Substituting $S_{ix,y,z}$ into the Hamiltonian (\ref{model}) and then do the Fourier transformation, we get the quadratic Hamiltonian in momentum space
\begin{align}
H=\frac{1}{2}\sum_{\bm{k}^\prime}\Psi_{\bm{k}^\prime}^\dagger h(\bm{k}^\prime)\Psi_{\bm{k}^\prime},
\end{align}
where $\Psi_{\bm{k}^\prime}^\dagger=(a_{1\bm{k}^\prime}^\dagger, a_{2\bm{k}^\prime}^\dagger, a_{3\bm{k}^\prime}^\dagger, a_{1-\bm{k}^\prime}, a_{2-\bm{k}^\prime}, a_{3-\bm{k}^\prime})$. The magnon Hamiltonian matrix is
\begin{align}
h(\bm{k}^\prime)=\left(
\begin{array}{cc}
A_{\bm{k}^\prime}&B_{\bm{k}^\prime}^\dagger\\[4pt]
B_{\bm{k}^\prime}&A_{\bm{k}^\prime}^*
\end{array}
\right)S,
\end{align}
with $A_{\bm{k}^\prime}$ and $B_{\bm{k}^\prime}$ $3\times3$ matrices. Their elements are as follows
\begin{align}
A_{11}=&-2J(c_z+e_z)-2K(c_8+e_7),\\
A_{22}=&-2J(c_z+d_z)-2K(c_8+d_6),\\
A_{33}=&-2J(d_z+e_z)-2K(d_6+e_7),\\
A_{12}=&[J(c_x+c_y-{\rm i}c_{xy}+{\rm i}c_{yx})+Kc_3]\cos \bm{k}^\prime\cdot\bm{\delta}_1,\\
A_{21}=&A_{12}^*,\\
A_{13}=&[J(e_x+e_y+{\rm i}e_{xy}-{\rm i}e_{yx})\notag\\
&+K(e_2+e_5+{\rm i}e_{10}-{\rm i}e_{12})]\cos \bm{k}^\prime\cdot\bm{\delta}_3,\\
A_{31}=&A_{13}^*,\\
A_{23}=&[J(d_x+d_y-{\rm i}d_{xy}+{\rm i}d_{yx})\notag\\
&+K(d_1+d_4-{\rm i}d_9+{\rm i}d_{11})]\cos \bm{k}^\prime\cdot\bm{\delta}_2,\\
A_{32}=&A_{23}^*,\\
B_{12}=&B_{21}=[J(c_x-c_y-{\rm i}c_{xy}-{\rm i}c_{yx})+Kc_3]\cos \bm{k}^\prime\cdot\bm{\delta}_1,\\
B_{13}=&B_{31}=[J(e_x-e_y-{\rm i}e_{xy}-{\rm i}e_{yx})\notag\\
&+K(e_2-e_5-{\rm i}e_{10}-{\rm i}e_{12})]\cos \bm{k}^\prime\cdot\bm{\delta}_3,\\
B_{23}=&B_{32}=[J(d_x-d_y-{\rm i}d_{xy}-{\rm i}d_{yx})\notag\\
&+K(d_1-d_4-{\rm i}d_9-{\rm i}d_{11})]\cos \bm{k}^\prime\cdot\bm{\delta}_2,\\
B_{11}=&B_{22}=B_{33}=0,
\end{align}
where
\begin{align}
c_1=&\cos\theta_1\cos\phi_1\cos\theta_2\cos\phi_2,\\
c_2=&\cos\theta_1\sin\phi_1\cos\theta_2\sin\phi_2,\\
c_3=&\sin\theta_1\sin\theta_2,\\
c_4=&\sin\phi_1\sin\phi_2,\\
c_5=&\cos\phi_1\cos\phi_2,\\
c_6=&\sin\theta_1\cos\phi_1\sin\theta_2\cos\phi_2,\\
c_7=&\sin\theta_1\sin\phi_1\sin\theta_2\sin\phi_2,\\
c_8=&\cos\theta_1\cos\theta_2,\\
c_9=&-\cos\theta_1\cos\phi_1\sin\phi_2,\\
c_{10}=&\cos\theta_1\sin\phi_1\cos\phi_2,\\
c_{11}=&-\sin\phi_1\cos\theta_2\cos\phi_2,\\
c_{12}=&\cos\phi_1\cos\theta_2\sin\phi_2,\\
c_x=&c_1+c_2+c_3,\\
c_y=&c_4+c_5,\\
c_z=&c_6+c_7+c_8,\\
c_{xy}=&c_9+c_{10},\\
c_{yx}=&c_{11}+c_{12},
\end{align}
and change the corresponding subscripts in $\theta_{1,2}$ and $\phi_{1,2}$ to $\theta_{2,3}$ and $\phi_{2,3}$, we get the corresponding expressions for $d_i$ with $i=1-12,x,y,z,xy,yx$. Similarly, change $\theta_{1,2}$ and $\phi_{1,2}$ to $\theta_{3,1}$ and $\phi_{3,1}$, we get the corresponding expressions for $e_i$. The vectors $\bm{\delta}_{1,2,3}$ are the NN vectors of the KL with $\bm{\delta}_1=(1/2,0)$, $\bm{\delta}_2=(-1,\sqrt{3})/4$, $\bm{\delta}_3=(-1,-\sqrt{3})/4$, which are defined in the new 2D frame $x^\prime y^\prime$. Note that to get the eigenvalues and eigenvectors of bosonic quadratic Hamiltonian, we need to diagonalize the matrix $I_-h(\bm{k}^\prime)$ instead of $h(\bm{k}^\prime)$, where
\begin{align}
I_-=\left(
\begin{array}{cc}
I&0\\[4pt]
0&-I
\end{array}
\right),
\end{align}
with $I$ the $3\times3$ identity matrix .

\bibliography{Reference}
%

\end{document}